\documentclass[conference]{IEEEtran}
\usepackage{cite}

\ifCLASSINFOpdf
   \usepackage[pdftex]{graphicx}
\else
\fi
\usepackage{amsmath}
\usepackage{slashed}
\usepackage{amsmath,amssymb,amsthm,verbatim,amsfonts,amscd,graphicx,float,physics}

\begin{document}
%
\title{Quantum Computation and Visualization of Hamiltonians Using Discrete Quantum Mechanics and IBM QISKit}

\author{\IEEEauthorblockN{Raffaele Miceli}
\IEEEauthorblockA{Physics Department\\
Stony Brook Univerity\\
Stony Brook, New York 11794\\
Email: raffaele.miceli.32@gmail.com}
\and
\IEEEauthorblockN{Michael McGuigan}
\IEEEauthorblockA{Computational Science Initiative\\
Brookhaven National Laboratory\\
Upton, New York 11973\\
Email: mcguigan@bnl.gov}}


%


\maketitle

\begin{abstract}
Quantum computers have the potential to transform the ways in which we tackle some important problems. The efforts by companies like Google, IBM and Microsoft to construct quantum computers have been making headlines for years. Equally important is the challenge of translating problems into a state that can be fed to these machines. Because quantum computers are in essence controllable quantum systems, the problems that most naturally map to them are those of quantum mechanics. Quantum chemistry has seen particular success in the form of the variational quantum eigensolver (VQE) algorithm, which is used to determine the ground state energy of molecular systems. The goal of our work has been to use the matrix formulation of quantum mechanics to translate other systems so that they can be run through this same algorithm. We describe two ways of accomplishing this using a position basis approach and a Gaussian basis approach. We also visualize the wave functions from the eigensolver and make comparisons to theoretical results obtained with continuous operators.
\end{abstract}


%
\IEEEpeerreviewmaketitle

\section{Introduction}

Once they are fully implemented, quantum computers will likely occupy a role similar to graphics cards in the overall computing landscape. Where graphics cards are very well suited to performing parallel computations, quantum computers can be used to perform calculations which span a vast amount of variables. The computational power of a quantum computer rises exponentially with its available qubits, outstripping supercomputers at a mere 50.

This project aims to leverage quantum computation to calculate the ground state energy of some Hamiltonians. In order to translate this program so that it can be fed to the quantum computer, we need to use the language of linear algebra. We will represent the operators of quantum mechanics with matrices and the states as vectors. To create and manipulate these objects I utilized Mathematica. For the quantum computation I utilized the QISKit Python library made by IBM. It allows us to load in the matrices we created and run quantum programs on them using either simulators or IBM's own quantum computers over the Cloud.

\section{Operators in the Position Basis}
In the position basis, we work on a lattice with $n$ sites indexed $\frac{1-n}{2},\frac{3-n}{2},\frac{5-n}{2},\dots,\frac{n-3}{2},\frac{n-1}{2}$. To make notation simpler, we introduce a mapping from the natural numbers to our lattice: 
\begin{equation} \label{eq:lattice}
\ell(a) = \frac{2a-1-n}{2}\ , \ a\ \in \ \{ 1,2,\dots,n\}
\end{equation}

The eigenvectors of our Hamiltonians will represent the probability density for a particle to be found on a particular site of the lattice.

We start by constructing the position operator as an $n \times n$ matrix:
\begin{equation} \label{eq:Xpos}
\bra{j}X_{pos}\ket{k} = \sqrt{\frac{2 \pi}{n}} \ \ell(j) \ \delta_{j,k}
\end{equation}

So the position operator is a traceless diagonal matrix made from our lattice indices with a scaling factor in front. To construct the momentum operator, we will first need a matrix for the discrete Fourier transform:

\begin{equation} \label{eq:Four}
\bra{j}F\ket{k} = \frac{1}{\sqrt{n}}e^{\frac{2 \pi i}{n} \ell(j,n) \ell(k,n)}
\end{equation}

Where $ \omega = e^{\frac{2 \pi i}{n}} $ is a primitive n-th root of unity. The exact form of the Fourier transform matrix depends on the lattice that we choose. The momentum operator is then

\begin{equation} \label{eq:Ppos}
P_{pos} = F^\dag X_{pos} F
\end{equation}

We proceeded in a fashion similar to the approach described by Jagannathan and Vasudevan in \cite{Jagannathan:1981rh}. In their paper they directly defined both their position and momentum operators and then detailed the relations between them and the finite Fourier transform. Instead we created the same position operator and Fourier transform and then used them to create a momentum operator that would satisfy all the same relations. We chose this path because it works for any size of matrix operator; the method described in the paper works only for odd values, and we need even values to implement these operators on a quantum computer, as we will explain later.

\section{Operators in the Harmonic Oscillator Energy Basis}

In the energy basis, the eigenfunctions of our Hamiltonians will represent the probability amplitudes for our system to have a certain energy. We start this time by constructing a discrete version of the annihilation operator as the $n \times n$ matrix

\begin{equation} \label{eq:AEn}
\bra{j}A\ket{k} = \sqrt{j} \ \delta_{j,k-1}
\end{equation}

Then our position and momentum operators are

\begin{equation} \label{eq:XPEn}
X_{en} = \frac{1}{\sqrt{2}}(A^\dag + A), \quad P_{en} = \frac{i}{\sqrt{2}}(A^\dag - A)
\end{equation}

where $i = \sqrt{\- 1}$.

Additional work on discrete quantum mechanics can be found in papers by Meurice \cite{Meurice:1991}, Singh et. al. \cite{Singh:2018qzk}, Macridin et. al. \cite{Macridin:2018gdw}, and Berenstein \cite{Berenstein:2018zif}.

\section{Building Hamiltonians}
Once we have made our operators in the desired basis, constructing our Hamiltonians proceeds as usual. The Hamiltonian of the quantum harmonic oscillator is written as

\begin{equation} \label{eq:nHho}
\hat{H}_{ho} = \frac{\hat{p}^2}{2m} + \frac{1}{2} m\omega^2\hat{x}^2
\end{equation}

Setting the constants $\hbar$, $\omega$, and $m$ to 1 and replacing the usual continuous operators with our discrete ones, we get

\begin{equation} \label{eq:Hho}
H_{ho} = \frac{P^2}{2} + \frac{X^2}{2}
\end{equation}

Similarly, the Hamiltonian for the anharmonic oscillator with a cubic term is given by

\begin{equation} \label{eq:Haho}
H_{aho} = \frac{P^2}{2} + \frac{X^2}{2} - \alpha X^3
\end{equation}

In general, Hamiltonians can be expressed as the sum of a kinetic term and a potential:

\begin{equation} \label{eq:genH}
H = \frac{P^2}{2} + V(X)
\end{equation}

In certain cases it is advantageous to construct our Hamiltonians from the raising and lowering operators instead. In this case we can use the inverses of equations \ref{eq:XPEn} to get these operators from the momentum and position operators:

\begin{equation} \label{eq:APos}
A = \frac{1}{\sqrt{2}}(X + iP), \quad A^\dag = \frac{1}{\sqrt{2}}(X - iP)
\end{equation}

Then our harmonic oscillator Hamiltonian, for example, can be written as:
\begin{equation} \label{eq:Hho2}
H_{ho} = A^\dag A + I/2
\end{equation}

Where $I$ is the $n \times n$ identity matrix.

\section{Comparison - Harmonic Oscillator}

We can compare the position and energy bases by implementing a few simple Hamiltonians in each and looking the spectra of the matrices. We start with the simple harmonic oscillator formulated using the position and momentum operators as in equation \ref{eq:Hho}.

\begin{figure}[H]
\centering
\begin{minipage}[b]{0.8\linewidth}
  \centering
  \includegraphics[width=\linewidth]{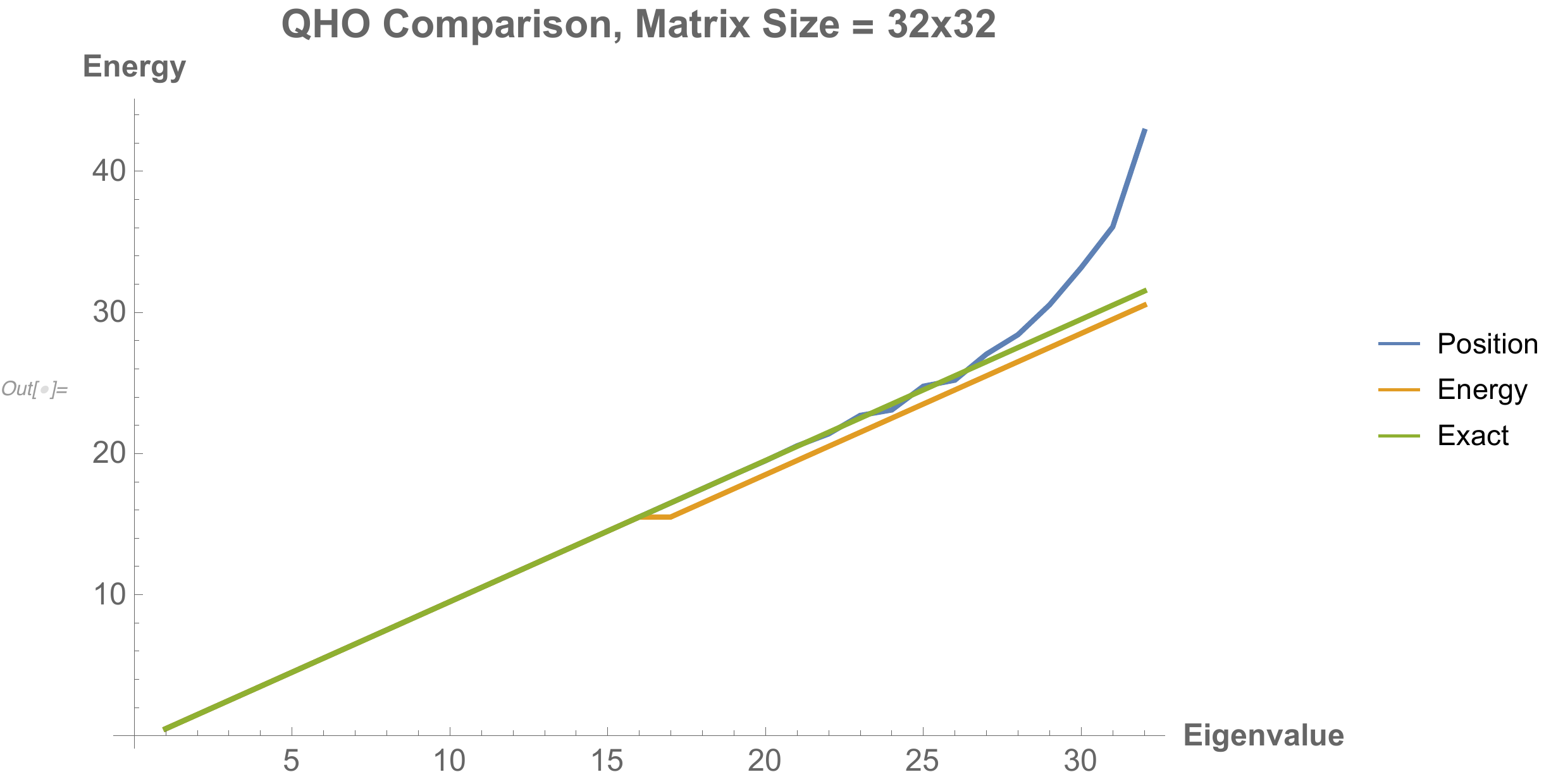}
  \vspace{3ex}
\end{minipage}
\caption{Comparison of energy and position bases for the quantum harmonic oscillator.}
\end{figure}

The energy levels of the harmonic oscillator are given by

\begin{equation}
E(n) = \hbar \omega \bigg( n + \frac{1}{2} \bigg) = n + \frac{1}{2}
\end{equation}

We observe that the position basis spectrum matches the exact expression very well for the first $3/4$ of the energies, but then diverges. The energy basis spectrum matches the exact expression, but has a kink in the middle. This happens because the Hamiltonian matrix in the energy basis has a degenerate eigenvalue. We could proceed by simply throwing out this ``junk'' eigenvalue, but when we get to more complicated Hamiltonians it could be more difficult to pick out. 

While playing around with the different bases, I noticed that if we instead express the Hamiltonian using the ladder operators as in equation \ref{eq:Hho2}, its spectrum matches the exact expression for any value of n. To investigate this, I tried substituting equations \ref{eq:XPEn} into equation \ref{eq:Hho} and expanding to get

\begin{equation} \label{eq:HhoA}
H_{ho} =\frac{1}{2}\big( A^\dag A + A A^\dag \big)
\end{equation}

The continuum ladder operators satisfy the commutation relation
\begin{equation} \label{eq:ladcom}
\left[ a,a^\dag \right] = 1
\end{equation}

If we take this to mean that our finite ladder operators should satisfy
\begin{equation}
\left[ A,A^\dag \right] = I
\end{equation}

then equation \ref{eq:HhoA} would match equation \ref{eq:Hho2} exactly. However, if we compute the commutator for the matrix ladder operators we find that it is

\begin{equation} \label{eq:ladcomM}
\left[ A,A^\dag \right] = 
\begin{pmatrix}
1 & 0 & 0 & \dots & 0 & 0 \\
0 & 1 & 0 & \dots & 0 & 0 \\
0 & 0 & 1 & \dots & 0 & 0 \\
\vdots & \vdots & \vdots & \ddots & \vdots & \vdots \\
0 & 0 & 0 &  \dots & 1 & 0 \\
0 & 0 & 0 &  \dots & 0 & -(n-1)
\end{pmatrix}
\end{equation}

This makes sense because matrix commutators are traceless. So what we have created here is a ``traceless identity'' matrix, which we'll call $\tilde{I}$. As the size of this matrix increases, it resembles the identity operator more and more. To get the spectrum to match our exact expression, we can use the expression

\begin{equation}
H_{ho}' = \frac{1}{2}(X^2 + P^2 - \tilde{I} + I) = A^\dag A + I/2
\end{equation}

Interestingly, if we try to use the form of equation \ref{eq:Hho2} with the position basis ladder operators, the spectrum will also have a (close to) duplicate eigenvalue in the middle.

\begin{figure}[H]
\centering
\begin{minipage}[b]{0.8\linewidth}
  \centering
  \includegraphics[width=\linewidth]{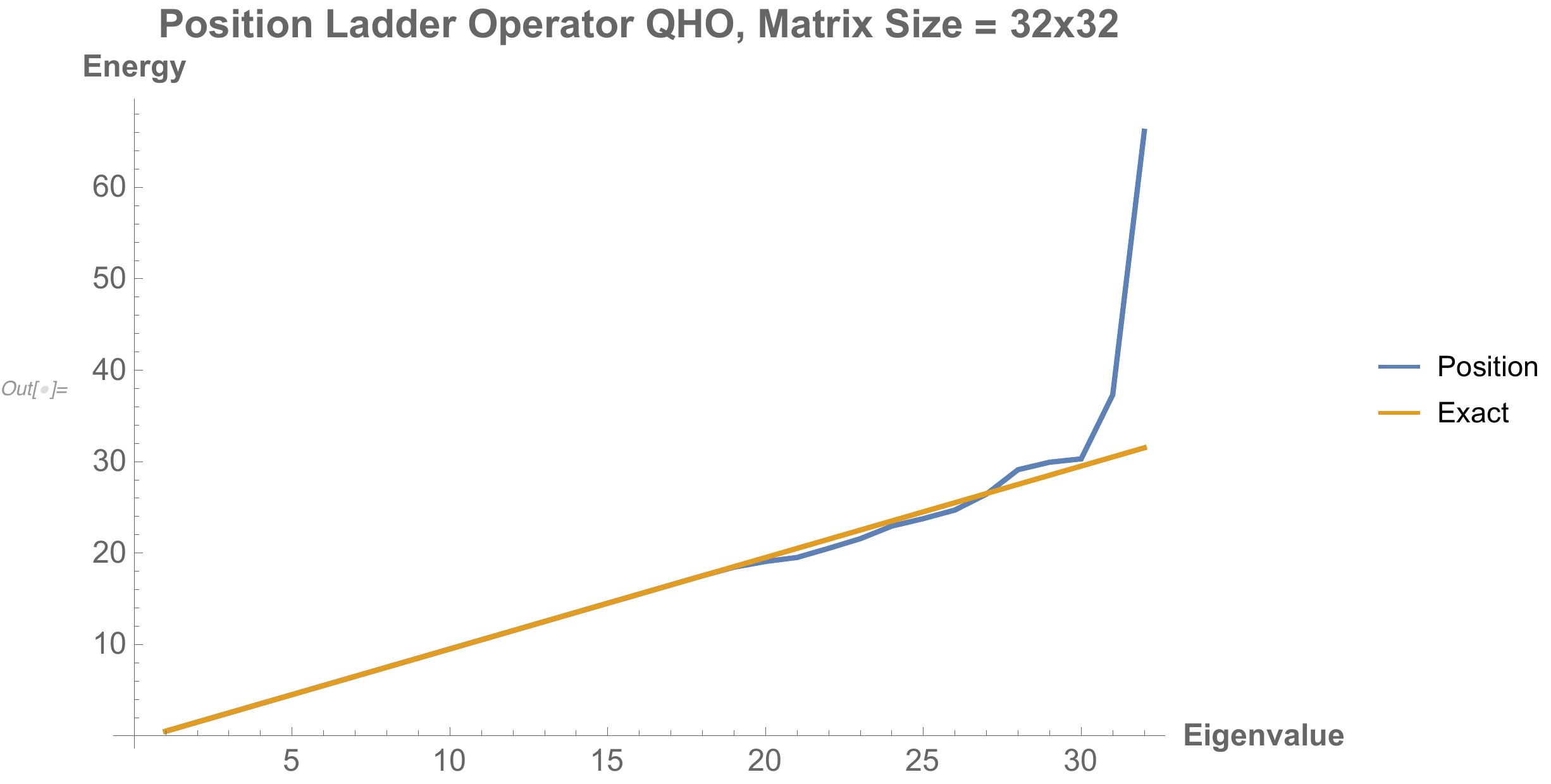}
\end{minipage}
\caption{Spectrum of QHO Hamiltonian made using position basis ladder operators, superimposed on exact energies}
\end{figure}

So it seems like we can intuit a tentative rule of thumb: If we want to work with ladder operators, we should use the energy basis; if we want to work with position and momentum operators, we should instead use the position basis.

Additional work on the discrete quantum harmonic oscillator can be found in papers by Atakishiyev et. al. \cite{Atakishiyev:2008}, Lorente \cite{Lorente:2001}, Barker et. al. \cite{Barker:2000},  and Aunola \cite{Aunola:2003}.

We also tested some perturbative Hamiltonians, namely the cubic and quartic anharmonic oscillators. We can compare the spectra of these systems to perturbative expressions formulated by Heisenberg. \cite{Heisenberg:1925} The potentials are:
\begin{equation} \label{eq:cub_aho}
V_{cub} = X^2 - \alpha X^3
\end{equation}
\begin{equation} \label{qua_aho}
V_{qua} = X^2 + \beta X^4 
\end{equation}

To implement these in the energy basis as just discussed, we simply append the perturbative part of each potential, written in terms of the position operator, to the basic quantum harmonic oscillator Hamiltonian written in terms of the ladder operators.

The perturbative energy expressions are:

\begin{equation} \label{eq:Wcub}
W_{cub}(n) = \hbar \omega_0 \bigg(n+\frac{1}{2}\bigg) - \frac{5 \lambda^2 \hbar^2}{12 m \omega_0^4} \bigg(n^2 +n + \frac{11}{30}\bigg)
\end{equation}

\begin{multline} \label{eq:Wquar}
W_{quar}(n) = \hbar \omega_0 \bigg(n+\frac{1}{2}\bigg)  + \frac{3 \lambda \hbar^2}{8 m \omega_0^2} \bigg(n^2 +n + \frac{1}{2}\bigg) \\ - \frac{\lambda^2 \hbar^2}{64 m^2 \omega_0^5} \bigg( 17 n^3 + \frac{51}{2}n^2 + \frac{59}{2}n +\frac{21}{2}\bigg)
\end{multline}

The following figures show a comparison between the energy basis, position basis, and perturbative expressions:

\begin{figure}[H]
\centering
\begin{minipage}[b]{0.8\linewidth}
  \centering
  \includegraphics[width=\linewidth]{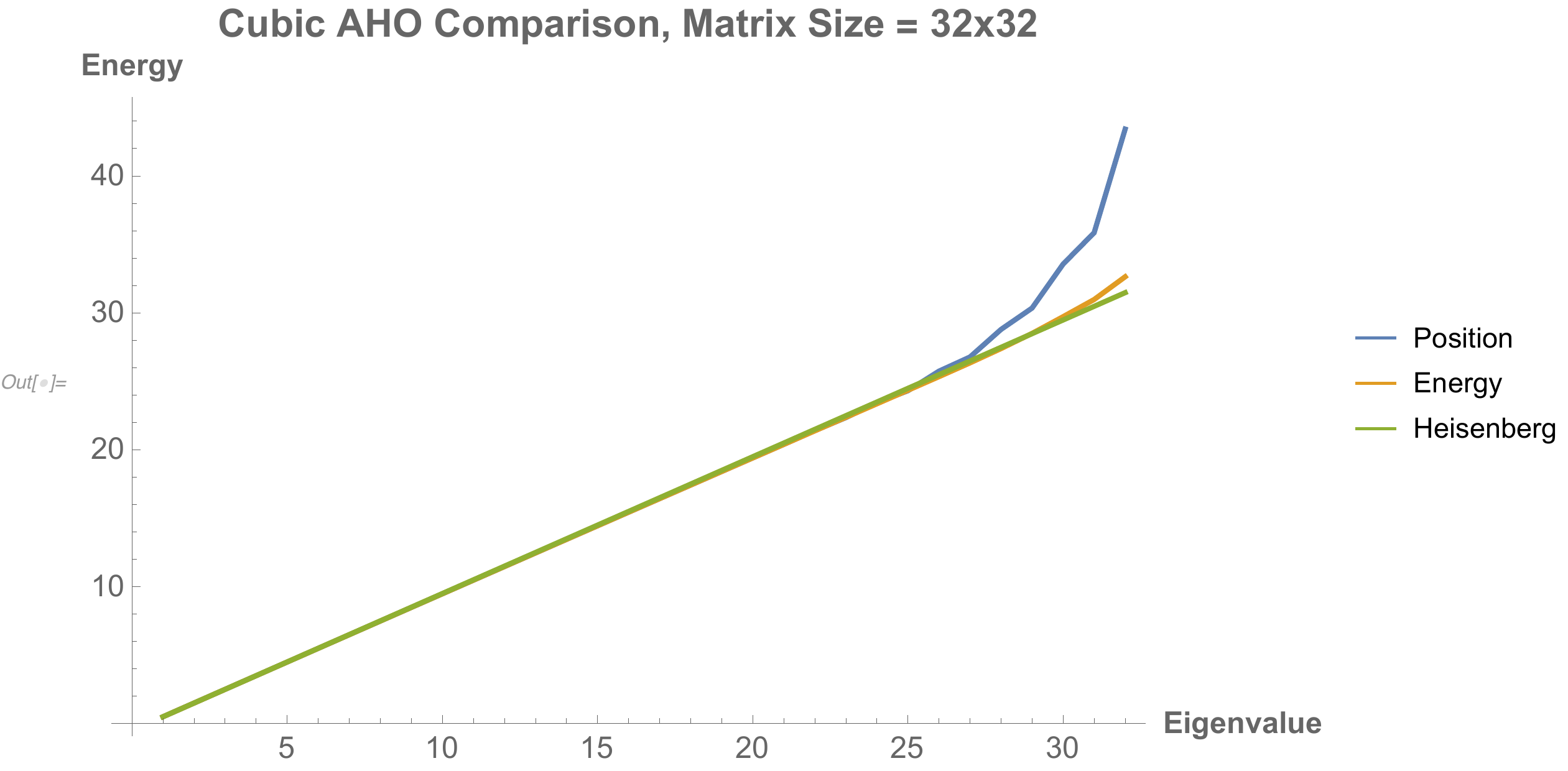}
  \vspace{3ex}
\end{minipage}
\caption{Comparison of energy and position bases for the cubic anharmonic oscillator with a coupling strength of 0.05.}
\end{figure}

\begin{figure}[H]
\centering
\begin{minipage}[b]{0.8\linewidth}
  \centering
  \includegraphics[width=\linewidth]{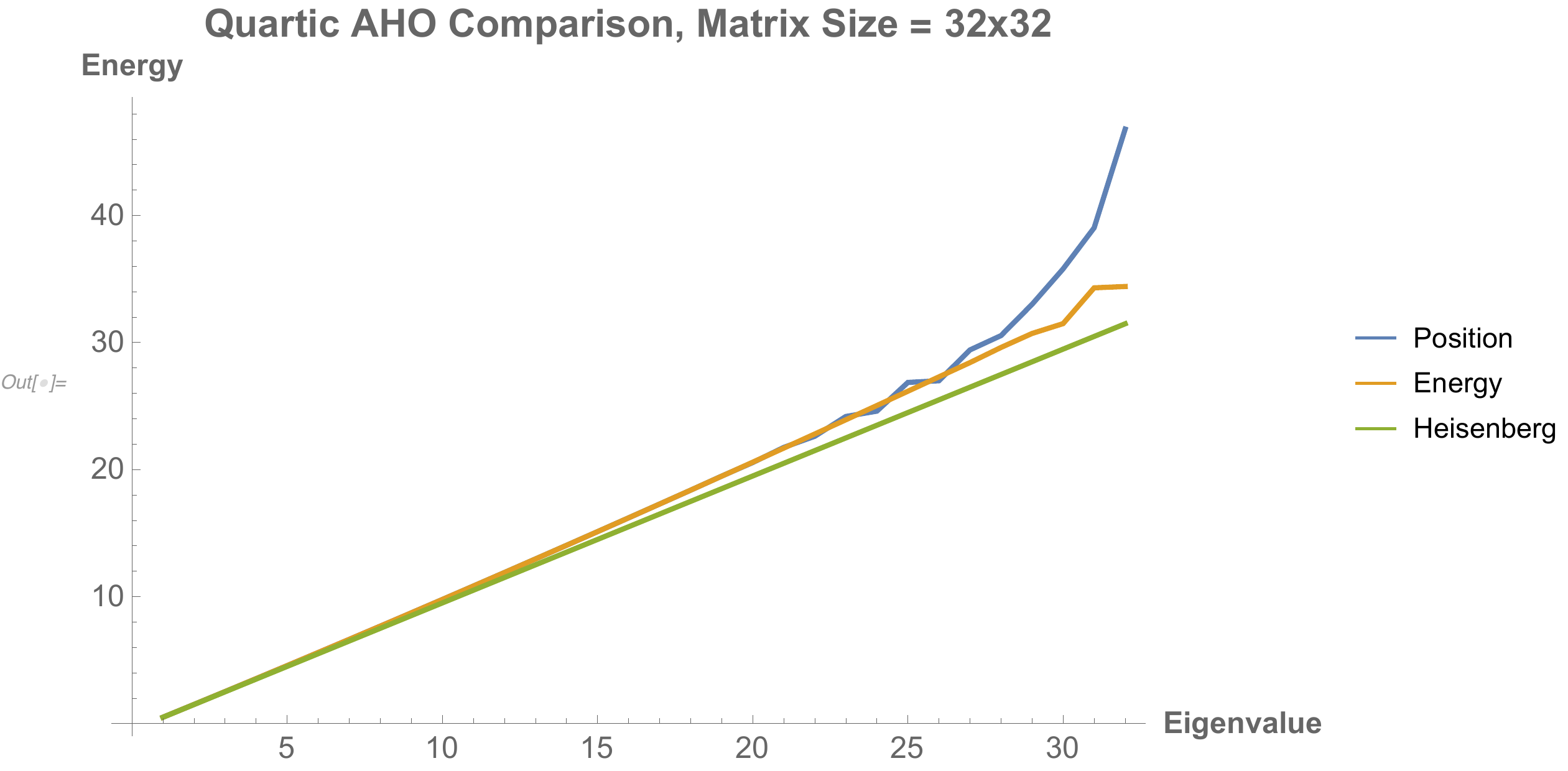}
  \vspace{3ex}
\end{minipage}
\caption{Comparison of energy and position bases for the quartic anharmonic oscillator with a coupling strength of 0.05.}
\end{figure}

In the cubic case, the two different spectra and the perturbative expression agree to within 1\% for over half the eigenvalues. In the quartic case, the agreement is a slightly worse 5\% for those same values. The agreement improves as we decrease the coupling to the perturbative component.

\section{Supersymmetric Oscillator}

Now we present another consistency check with a more complicated potential, namely the supersymmetric anharmonic oscillator. For the normal harmonic oscillator we use bosonic operators. Since bosons can exist in an infinite number of excited states, their matrix operators can be extended indefinitely. On the other hand, fermions can only be in two states, which we'll call $\ket{0}$ and $\ket{1}$. The lowering and raising operators, $C$ and $C^\dag$ respectively, must obey the following relations:
\begin{gather} \label{eq:flad}
C\ket{0} = 0, \quad C\ket{1} = \ket{0}; \\
C^\dag\ket{0} = \ket{1}, \quad C^\dag\ket{1} = 0
\end{gather}

Whereas with bosons the only ``illegal'' operation is lowering the ground state, with fermions you can't raise the single excited state either. These conditions are satisfied by the $2 \times 2$ matrices:

\begin{equation} \label{eq:fermcs}
C = 
\begin{pmatrix}
0 & 1 \\ 0 & 0
\end{pmatrix}, \quad
C^\dag = 
\begin{pmatrix}
0 & 0 \\ 1 & 0
\end{pmatrix}
\end{equation}

The state vectors for fermions are then simply:

\begin{equation} \label{eq:fermvecs}
\ket{0} = 
\begin{pmatrix}
1 \\ 0
\end{pmatrix}, \quad
\ket{1} = 
\begin{pmatrix}
0 \\ 1
\end{pmatrix}
\end{equation}

In the simplest supersymmetric system, we have a boson paired with a fermion. In the context of matrix operators, this pairing is achieved by extending our original operators using Kronecker products. \cite{Woit:2017vqo} For a bosonic operator:
\begin{equation} \label{eq:bosop}
\tilde{O}_B = O_B\ \otimes \ I_F
\end{equation}

And for a fermionic operator:
\begin{equation} \label{eq:fermop}
\tilde{O}_F = I_B \ \otimes \ O_F
\end{equation}

Here $I_b$ and $I_f$ are the identity matrices that are the same size as our bosonic and fermionic spaces, respectively. From now on we will use $A$, $C$, etc. to refer to the new extended operators. We will work with extended energy basis ladder operators, and the other operators can be built from these as described before.

Yu Musin \cite{Musin:1990sq} uses the following Hamiltonian:
\begin{gather} \label{eqe:musinH}
H = H_0 + g(U_0 + q \left[ a^\dag,a \right]); \\
H_0 = \frac{1}{2}\big(p^2 + \omega_0 q^2 + \omega_0 \left[ a^\dag,a \right]\big); \\
\quad U_0 = \omega_0 q^3 + \frac{1}{2} g q^4
\end{gather}

Here $q$ and $p$ are the position and momentum operators, $\omega_0$ and $g$ are constants, and $a$ and $a^\dag$ are actually the fermionic ladder operators. $H_0$ is the supersymmetric harmonic oscillator and the rest is the anharmonic perturbation.  Expanding, rewriting in terms of our matrix operators and replacing the non-perturbative part with ladder operators:
\begin{multline} \label{eq:musinH2}
H = \omega_0\Big(A^\dag A + \frac{1}{2} I\Big)  \\ +\frac{1}{2}\Big(2 \omega_0 g X^3 + g^2 X^4 + \big(\omega_0 I + 2 g X \big) \left[ C^\dag , C\right] \Big)
\end{multline}

Musin also calculated a perturbative expression for the energy levels of this Hamiltonian, which depends on $n_B$ and $n_F$, respectively the energy levels of our boson and fermion\footnote{There was a typo in the original paper: the $g$ in the second term should have been $g^2$. Equation \ref{eq:musinE} reflects this change.}:
\begin{multline} \label{eq:musinE}
E_n = \hbar\omega_0\big(n_B + n_F) + \frac{3\hbar^2}{4\omega_0^2}g^2\Big(n_B^2 + n_B + \frac{1}{2}\Big) - \\ \frac{15\hbar^2}{4\omega_0^2}g^2\Big(n_B^2 + n_B + \frac{11}{30}\Big)
\end{multline}

When we compare the spectrum of equation \ref{eq:musinH2} with the corresponding values of equation \ref{eq:musinE}, we see again that they match quite closely for the first half of the eigenvalues:

\begin{figure}[H]
\centering
\begin{minipage}[b]{0.8\linewidth}
  \centering
  \includegraphics[width=\linewidth]{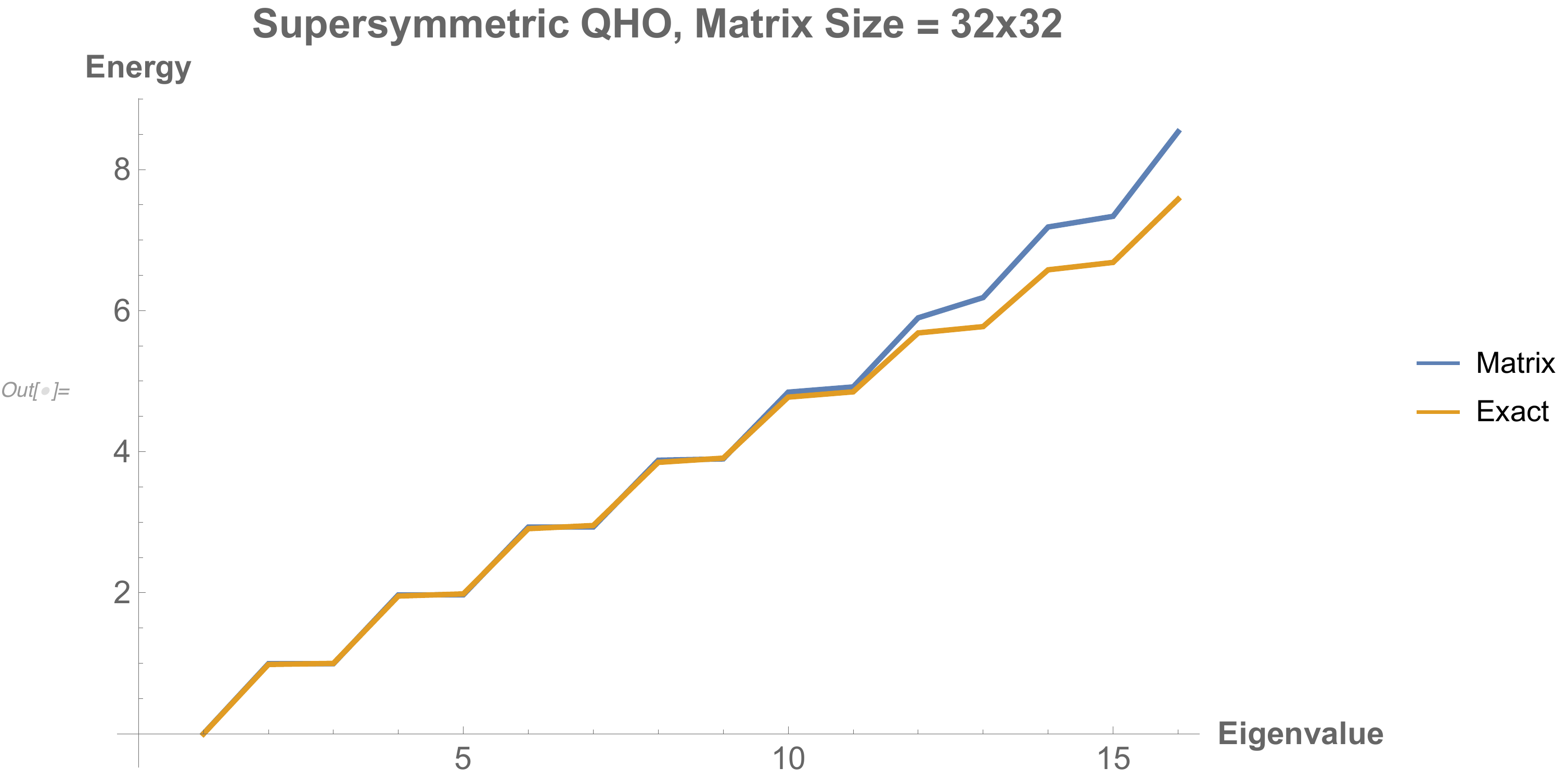}
  \vspace{3ex}
\end{minipage}
\caption{Comparison of supersymmetric matrix Hamiltonian and perturbative expression from \cite{Musin:1990sq}, with a coupling strength of 0.05.}
\end{figure}

\section{Quantum Computing}

\subsection{Brief Overview}

The basic operational units of quantum computers are qubits. Qubits in turn are two-state quantum systems, be they atomic spins or photon polarizations. The quantum computer consists of the qubits, the connections between them, and the apparatus used to manipulate their states. In most cases the qubits are very sensitive to mechanical and thermal perturbations, which can cause them to decohere and lose their quantum properties. To prevent this from happening the whole machine is encased in a multistage dilution refrigerator to keep it cold and housed in a shock absorbent chamber to keep it very still. Even with all of these precautions, qubits still typically decohere in a matter of microseconds.


Calculations are performed on a quantum computer by acting on single or multiple qubits via quantum logic gates and thereby changing their states. Since these are \textit{quantum} systems, they can exhibit superposition and entanglement, which are key to their computational power. When a qubit is in superposition it has a probability of being measured in either of its two states. Entanglement refers to a state involving two qubits which can't be described independently. If we measure one of them, we automatically know the state of the other.

We can conceptualize a qubit as a complex two-component vector:

\begin{equation} \label{eq:qubit}
\ket{q} = 
\begin{pmatrix}
\alpha \\ \beta
\end{pmatrix} = \alpha \ket{0} + \beta \ket{1}, \quad
||\alpha||^2 + ||\beta||^2 = 1
\end{equation}

Here $\alpha$ and $\beta$ represent the probability amplitudes for the qubit being in its two states, so the sum of their squares must be one. With this in hand, we can think of single-qubit quantum gates as $2 \times 2$ unitary matrices. They must be unitary to conserve probability. This formalism also allows us to think of the state of a qubit as existing on the surface of a unit sphere called the Bloch sphere:

\begin{figure}[H]
\centering
\includegraphics[width=.3\linewidth]{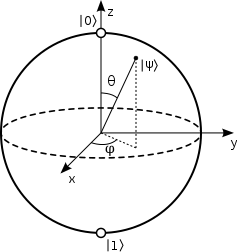}  
\caption{Drawing of the Bloch sphere. The two basis states of the qubit are represented by the poles of the sphere. (Image credit: Wikipedia)}
\end{figure}

Using the two angles $\theta$ and $\phi$, we can express $\alpha$ and $\beta$ as

\begin{equation} \label{eq:albet}
\alpha = \cos{\frac{\theta}{2}}, \quad \beta = e^{i\phi}\sin{\frac{\theta}{2}}
\end{equation}

So quantum logic gates simply move the point representing the state of the qubit along the Bloch sphere. Some single-qubit gates include rotations about the x-, y- and z-axes, the familiar Pauli matrices, and the Hadamard gate which can put a qubit into superposition. Arguably the most important 2-qubit gate is the controlled-not or ``CNOT'' gate, which is commonly used to entangle two neighboring qubits.

Quantum gates are used to build up quantum circuits, which take our set of qubits to a desired state. Typically a circuit will end with all the qubits being measured to get the result. Since we are dealing with probabilistic objects, we can't get an ``answer'' from a single measurement; we need to run the circuit many times and histogram all the results to get a probability distribution. The peaks in the distribution tell us what the outcome of the quantum program.

\subsection{The Variational Quantum Eigensolver (VQE) Algorithm}

The VQE algorithm \cite{Peruzzo:2013nat} tries to approximate the ground state energy of a given matrix Hamiltonian through a hybrid classical/quantum optimization process. We start by initializing our qubits with a trial state populated with guessed parameters. The quantum computer then evaluates the expectation value of this state with the given Hamiltonian. Using the returned value, a classical optimization algorithm varies the parameters. These two steps cycle until a user-set goal is set, either for number of steps or stability of the calculated energy.

In actuality, the algorithm is not run on the whole Hamiltonian at once, since in most cases it can't be represented as a single layer of quantum gates. To be fed into the VQE algorithm the Hamiltonian matrix must be decomposed into a sum of tensor products of Pauli matrices. This can be done because the Pauli matrices and the identity matrix form a complete basis for complex $2x2$ matrices, and their complete set of tensor products does the same for larger matrices.

\subsection{Scripts}

My workflow included three major steps:

\begin{enumerate}
\item Create Hamiltonian matrix (Mathematica)
\item Decompose Hamiltonian matrix into sum of Pauli tensor products (Mathematica)
\item Run VQE algorithm (IBM QISKit in Python)
\end{enumerate}

If we limit ourselves to a small quantity of qubits - no more than 10 or so - we can use Mathematica to build the operators and make Hamiltonians with them. For each different type of Hamiltonian I wrote a script that would build the matrix and export it to a file as an array. I then wrote a second script that would take the matrix file as an input and output a list of Pauli tensor products and their corresponding coefficients. 

The second script works by generating a list of all possible Pauli matrix tensor product for a given number of qubits, then taking the trace of the product of each with the Hamiltonian. This trace, multiplied by a normalization factor, is the coefficient we want. Each Pauli tensor with a non-zero coefficient is then written to an output file. It is this file which is read by the VQE script.

The VQE script's main inputs are listed below:

\begin{enumerate}
\item Hamiltonian matrix as list of Pauli tensors in file
\item variational form (trail circuit) defined in separate Python script
\item quantum depth (number of times to repeat circuit in a single cycle)
\item maximum number of optimization cycles
\item device to use, either local simulator or IBM's devices via the Cloud
\end{enumerate}

Once it runs, the VQE script outputs a convergence plot, a circuit diagram depicting the trial function, and a log file. The outputs of runs on the same type of Hamiltonian are automatically saved to their own directory, and the logs of different runs are appended to the same file. All the runs I've done so far have been on a local simulator instead of IBM's real quantum devices. For example, a run of a 6-cubit anharmonic oscillator Hamiltonian with a simple trial function, quantum depth of 3 and just 100 optimization steps took about 45 minutes to finish and did not converge completely to the exact energy.

We saw from limited runs that using the local simulator will require high-performance computing (HPC) for large runs. The harmonic and anharmonic oscillator Hamiltonian runs converged correctly when using a very simple circuit. More complicated Hamiltonians, involving supersymmetry and/or finite temperature, will require additional study. Example outputs for the harmonic oscillator, showing a convergence to within 2\% of the expected ground state energy, can be seen in figure \ref{fig:results} below.

\begin{figure}[H]
\centering
\begin{minipage}{.7\linewidth}
  \centering
  \includegraphics[width=\linewidth]{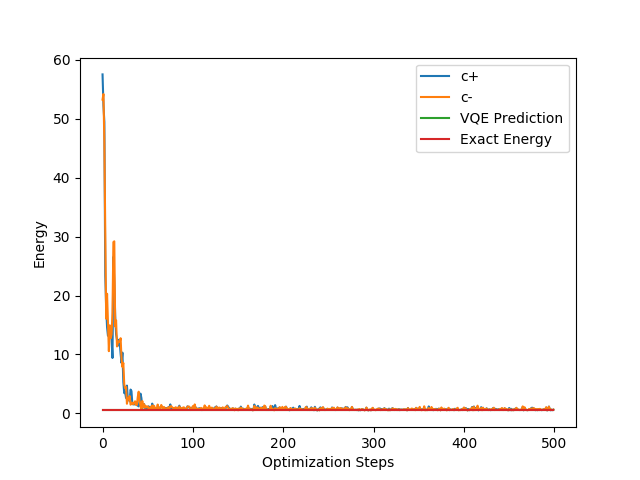}
\vspace{3ex}
\end{minipage}
\begin{minipage}{.7\linewidth}
  \centering
  \includegraphics[width=\linewidth]{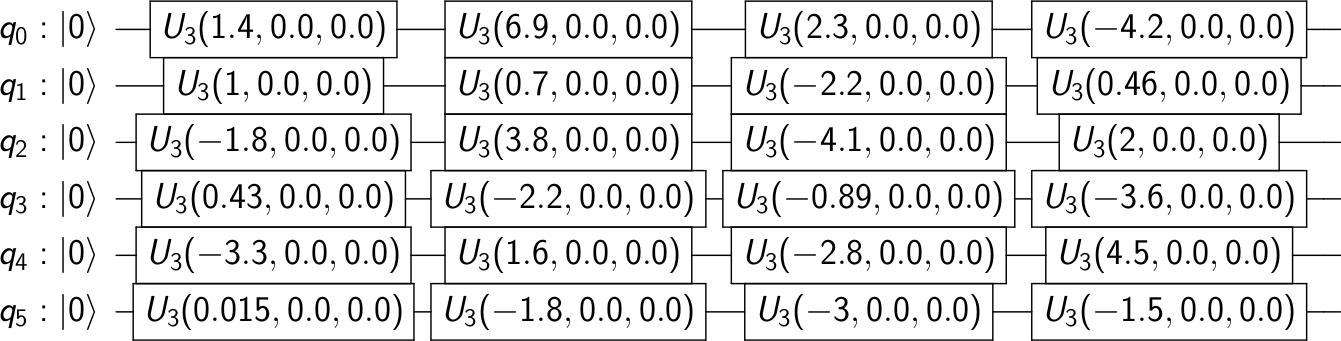}
\vspace{3ex}
\end{minipage}
\begin{minipage}{.6\linewidth}
  \centering
  \includegraphics[width=\linewidth]{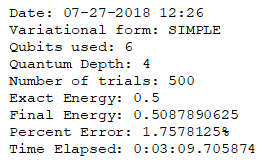}
\end{minipage}
\caption{Examples of the outputs from a VQE run. \textbf{Top left}: a convergence plot showing the two different paths taken by the optimization algorithm; \textbf{Top right}: a quantum circuit diagram with optimized parameters; \textbf{Bottom}: the log entry for the run.}
\label{fig:results}
\end{figure}

For additional information on discrete quantum mechanics using quantum computers, see the recent paper by Klco et. al. \cite{Klco:2018zqz}.

\section{Conclusion}

Though the results are still rather fresh, we've seen some success running simple Hamiltonians on quantum computers using simple trial functions. Moving to bigger and/or more complicated Hamiltonians presents various challenges. Due to the nature of the tensor product, a composite system with just a few simple parts could have a monstrously large matrix Hamiltonian. Making larger matrices will likely become too time-consuming for Mathematica, so we'll have to migrate our scripts to a lower level programming language. This would still leave us with the issue of efficiently storing and accessing these matrices, which will quickly outgrow the space on a typical laptop. The greatest challenge in fully exploiting the VQE algorithm will be designing good trial functions, which often need to be tailored to specific Hamiltonians. Even if we find one that works well for a given Hamiltonian, its effectiveness will not necessarily persist as we scale up the matrix size. We have also been working on implementing a Hamiltonian that describe finite temperature effects on a supersymmetric system, based on work by Das et al \cite{Das:1989cj}, as well as a matrix effective model studied by Kashiwa et. al. \cite{Kashiwa:2012wa}. The former proved difficult to implement because of the complexity of the theory and the fact that the size of the matrix Hamiltonian grows more rapidly than usual as we increase the size of the boson components.

\section{Acknowledgements}

This project was supported in part by the Brookhaven National Laboratory (BNL), Computational Science Initiative under the BNL Supplemental Undergraduate Research Program (SURP) . The author would like to thank Michael McGuigan for many stimulating discussions, as well as past and present CSI interns for their help setting up the Mathematica and Python code. We acknowledge use of the IBM Q for this work. The views expressed are those of the authors and do not reflect the official policy or position of IBM or the IBM Q team. Michael McGuigan is supported from DOE HEP Office of Science DE-SC0019139: Foundations of Quantum Computing for Gauge Theories and Quantum Gravity.

\bibliographystyle{IEEEtran}
\bibliography{references}

\begin{thebibliography}{10}
\providecommand{\url}[1]{#1}
\csname url@samestyle\endcsname
\providecommand{\newblock}{\relax}
\providecommand{\bibinfo}[2]{#2}
\providecommand{\BIBentrySTDinterwordspacing}{\spaceskip=0pt\relax}
\providecommand{\BIBentryALTinterwordstretchfactor}{4}
\providecommand{\BIBentryALTinterwordspacing}{\spaceskip=\fontdimen2\font plus
\BIBentryALTinterwordstretchfactor\fontdimen3\font minus
  \fontdimen4\font\relax}
\providecommand{\BIBforeignlanguage}[2]{{%
\expandafter\ifx\csname l@#1\endcsname\relax
\typeout{** WARNING: IEEEtran.bst: No hyphenation pattern has been}%
\typeout{** loaded for the language `#1'. Using the pattern for}%
\typeout{** the default language instead.}%
\else
\language=\csname l@#1\endcsname
\fi
#2}}
\providecommand{\BIBdecl}{\relax}
\BIBdecl

\bibitem{Jagannathan:1981rh}
R.~Jagannathan, T.~S. Santhanam, and R.~Vasudevan, ``{Finite Dimensional
  Quantum Mechanics of a Particle},'' \emph{Int. J. Theor. Phys.}, vol.~20, p.
  755, 1981.

\bibitem{Meurice:1991}
\BIBentryALTinterwordspacing
Y.~Meurice, ``A discretization of p-adic quantum mechanics,''
  \emph{Communications in Mathematical Physics}, vol. 135, no.~2, pp. 303--312,
  Jan 1991. [Online]. Available: \url{https://doi.org/10.1007/BF02098045}
\BIBentrySTDinterwordspacing

\bibitem{Singh:2018qzk}
A.~Singh and S.~M. Carroll, ``{Modeling Position and Momentum in
  Finite-Dimensional Hilbert Spaces via Generalized Clifford Algebra
  1806.10134},'' 2018.

\bibitem{Macridin:2018gdw}
A.~Macridin, P.~Spentzouris, J.~Amundson, and R.~Harnik, ``{Electron-Phonon
  Systems on a Universal Quantum Computer},'' \emph{Phys. Rev. Lett.}, vol.
  121, no.~11, p. 110504, 2018.

\bibitem{Berenstein:2018zif}
D.~Berenstein, ``{A toy model for time evolving QFT on a lattice with
  controllable chaos 1803.02396},'' 2018.

\bibitem{Atakishiyev:2008}
N.~M. {Atakishiyev}, A.~U. {Klimyk}, and K.~B. {Wolf}, ``{A discrete quantum
  model of the harmonic oscillator},'' \emph{Journal of Physics A Mathematical
  General}, vol.~41, no.~8, p. 085201, Feb. 2008.

\bibitem{Lorente:2001}
\BIBentryALTinterwordspacing
M.~Lorente, ``Continuous vs. discrete models for the quantum harmonic
  oscillator and the hydrogen atom,'' \emph{Physics Letters A}, vol. 285,
  no.~3, pp. 119 -- 126, 2001. [Online]. Available:
  \url{http://www.sciencedirect.com/science/article/pii/S0375960101003036}
\BIBentrySTDinterwordspacing

\bibitem{Barker:2000}
\BIBentryALTinterwordspacing
L.~Barker, Ãagatay Candan, T.~Hakioglu, M.~A. Kutay, and H.~M. Ozaktas,
  ``The discrete harmonic oscillator, harper's equation, and the discrete
  fractional fourier transform,'' \emph{Journal of Physics A: Mathematical and
  General}, vol.~33, no.~11, p. 2209, 2000. [Online]. Available:
  \url{http://stacks.iop.org/0305-4470/33/i=11/a=304}
\BIBentrySTDinterwordspacing

\bibitem{Aunola:2003}
\BIBentryALTinterwordspacing
M.~Aunola, ``The discretized harmonic oscillator: Mathieu functions and a new
  class of generalized hermite polynomials,'' \emph{Journal of Mathematical
  Physics}, vol.~44, no.~5, pp. 1913--1936, 2003. [Online]. Available:
  \url{https://doi.org/10.1063/1.1561156}
\BIBentrySTDinterwordspacing

\bibitem{Heisenberg:1925}
\BIBentryALTinterwordspacing
W.~Heisenberg, ``{\"U}ber quantentheoretische umdeutung kinematischer und
  mechanischer beziehungen.'' \emph{Zeitschrift f{\"u}r Physik}, vol.~33,
  no.~1, pp. 879--893, Dec 1925. [Online]. Available:
  \url{https://doi.org/10.1007/BF01328377}
\BIBentrySTDinterwordspacing

\bibitem{Woit:2017vqo}
\BIBentryALTinterwordspacing
P.~Woit, \emph{{Quantum Theory, Groups and Representations}}.\hskip 1em plus
  0.5em minus 0.4em\relax Springer, 2017. [Online]. Available:
  \url{http://inference-review.com/article/woits-way}
\BIBentrySTDinterwordspacing

\bibitem{Musin:1990sq}
{\relax Yu}.~R. Musin, ``{A Supersymmetric anharmonic oscillator},'' \emph{Sov.
  Phys. J.}, vol.~33, pp. 401--404, 1990.

\bibitem{Peruzzo:2013nat}
A.~Peruzzo, J.~McClean, P.~Shadbolt, M.~Yung, X.~Zhou, P.~J. Love,
  A.~Aspuru-Guzik, and J.~L. O'Brien, ``A variational eigenvalue solver on a
  quantum processor,'' 2013.

\bibitem{Klco:2018zqz}
N.~Klco and M.~J. Savage, ``{Digitization of Scalar Fields for NISQ-Era Quantum
  Computing 1808.10378},'' 2018.

\bibitem{Das:1989cj}
A.~K. Das, ``{Supersymmetry and Finite Temperature},'' \emph{Physica}, vol.
  A158, pp. 1--21, 1989.

\bibitem{Kashiwa:2012wa}
K.~Kashiwa, R.~D. Pisarski, and V.~V. Skokov, ``{Critical endpoint for
  deconfinement in matrix and other effective models},'' \emph{Phys. Rev.},
  vol. D85, p. 114029, 2012.

\end{thebibliography}

\end{document}